\documentclass[journal,letterpaper,twocolumn]{IEEEtran}
\usepackage{amsfonts}
\usepackage{amssymb}
\usepackage{cite}
\usepackage[cmex10]{amsmath}
\usepackage{float}
\usepackage{color}
\usepackage{stfloats,fancyhdr}

\usepackage{algorithm}
\usepackage{algorithmic}
\usepackage{multirow}
\usepackage{changepage}
\usepackage[normalem]{ulem}

\newtheorem{remark}{Remark}

\IEEEoverridecommandlockouts

\ifCLASSINFOpdf
  \usepackage[pdftex]{graphicx}
  \DeclareGraphicsExtensions{.pdf,.jpeg,.png}
\else
  \usepackage[dvips]{graphicx}
  \DeclareGraphicsExtensions{.eps}
\fi

\usepackage{subfigure}

\usepackage{fancybox,dashbox}
\begin{document}

\title{Ultra-Reliable Short-Packet Communications: Half-Duplex or Full-Duplex Relaying?
\thanks{Y. Gu, H. Chen, Y. Li and B. Vucetic are with School of Electrical and Information Engineering, The University of Sydney, Sydney, NSW 2006, Australia (email: yifan.gu@sydney.edu.au, he.chen@sydney.edu.au, yonghui.li@sydney.edu.au, branka.vucetic@sydney.edu.au).}
%\thanks{H. Chen, Y. Li, and B. Vucetic are with School of Electrical and Information Engineering, The University of Sydney, Sydney, NSW 2006, Australia (email: he.chen@sydney.edu.au, yonghui.li@sydney.edu.au, branka.vucetic@sydney.edu.au).}
}

\author{Yifan Gu, He Chen, Yonghui Li
%\IEEEmembership{Student Member,~IEEE,}
%,~\IEEEmembership{Member,~IEEE,}
%%Yonghui Li,~\IEEEmembership{Senior Member,~IEEE,} \\
%%Raymond~H.~Y.~Louie,~\IEEEmembership{Member,~IEEE,}
 and Branka Vucetic}

%\author{\IEEEauthorblockN {He (Henry) Chen\IEEEauthorrefmark{1}, %Xiangyun Zhou\IEEEauthorrefmark{2},
%}
%%\IEEEauthorblockA{\IEEEauthorrefmark{1}School of
%%Electrical and Information Engineering, The University of Sydney\\
%%Sydney, Australia, Email: he.chen@sydney.edu.au}
%%\IEEEauthorblockA{\IEEEauthorrefmark{2}Research School of Engineering, The Australian National University\\ Canberra, Australia, Email: xiangyun.zhou@sydney.edu.au}
%%\IEEEauthorblockA{\IEEEauthorrefmark{3}State Key Lab. of Mobile
%%Communications, Southeast University\\
%%Nanjing, China, 210096, Email: hongjixu@sdu.edu.cn}
%}

%\markboth{IEEE TRANSACTIONS ON WIRELESS COMMUNICATIONS
%%, VOL. XX, NO. XX, XXXX 2010
%}{He Chen \MakeLowercase{\textit{et al.}}:  }

\maketitle

\begin{abstract}
\iffalse
In this letter, we revisit the performance comparison of full-duplex relaying (FDR) and half-duplex relaying (HDR) under the finite blocklength regime. Specifically, we consider a classical three-node communication network consisting of one single-antenna source, one single-antenna destination and one relay equipped with isolated transmit and receive antennas. We derive approximate closed-form expressions of the block error rate (BLER) for both FDR and HDR systems using finite blocklength coding. Simple asymptotic expressions at high signal-to-noise ratio (SNR) for both systems are also derived. Based on the asymptotic expressions, we first optimize the transmit powers of source and relay subject to individual peak power constraints. We then compare the minimum blocklength (i.e., delay) of FDR and HDR systems with optimal transmit power under a given BLER constraint. We define and attain a closed-form expression of a critical BLER, which can be used to choose the optimal duplex mode efficiently for short-packet communication scenarios. Numerical results are provided to validate all the theoretical analysis. It is shown that FDR is more appealing to the system with relatively lower transmit power constraint, less stringent BLER requirement and stronger loop interference suppression.
\fi
This letter analyzes and compares the performance of full-duplex relaying (FDR) and half-duplex relaying (HDR) for ultra-reliable short-packet communications. Specifically, we derive both approximate and asymptotic closed-form expressions of the block error rate (BLER) for FDR and HDR using short packets with finite blocklength codes. We define and attain a closed-form expression of a critical BLER, which can be used to efficiently determine the optimal duplex mode for ultra-reliable low latency communication scenarios. Our results unveil that FDR is more appealing to the system with relatively lower transmit power constraint, less stringent BLER requirement and stronger loop interference suppression.
\end{abstract}


\IEEEpeerreviewmaketitle

\section{Introduction}
\iffalse
Compared to the current broadband services which are mainly designed for human-operated terminals, the majority of wireless connections in the fifth-generation (5G) cellular systems will be autonomous machines, referred to machine to machine communication (M2M) \cite{background1}. M2M requires efficient short-packet communications with stringent latency and reliability requirements. Very recently, the concept of ultra-reliable low latency communication (URLLC) has been proposed to address these new challenges in 5G M2M systems. URLLC refers to communication services where short-packet data is exchanged with relatively low throughput but stringent latency and reliability requirements \cite{towards}.
\fi

In conventional wireless systems, extremely long packet has been used in coding and transmission. However, to support ultra-reliable low latency communications (URLLC) as a new service type of 5G cellular systems \cite{henryref}, short-packet transmissions are essential. In this case, errors cannot be reduced to arbitrarily low for a given coding rate due to the limited packet size. Motivated by this, \cite{FBLtheory1} developed a new fundamental framework for short-packet communications and derived an error probability bound for a given blocklength and coding rate. It was shown that the block error rate (BLER) increases as the blocklength of the system decreases. This new theoretical framework opens a new research direction for the revisit of conventional communication networks, which were mainly designed based on the Shannon formula and thus cannot be directly applied to short-packet communications.

References \cite{yulin1,yulin3} investigated a classical three node cooperative network under finite blocklenth regime with static and quasi-static channel conditions. It was shown that there exists a performance tradeoff between cooperative and non-cooperative communications: On the one hand, a relay can boost the power gains of both hops and thus improve the system performance; on the other hand, it can degrade the system performance by halving the blocklength of each packet transmission. However, only half-duplex relaying (HDR) was considered in \cite{yulin1,yulin3}. To the best knowledge of the authors, the performance of full-duplex relaying (FDR) under the finite blocklength regime has not been studied so far in open literature. Compared to HDR, FDR has a great potential to improve the system performance since it has a longer blocklength for each packet transmission, which is twice longer as that of HDR. On the other side, FDR suffers from additional interference caused by receiving and transmitting information at the same time, degrading the system performance \cite{Full_duplex_outage, Full_duplex_adaption}. It is thus essential to revisit the optimal choice between FDR and HDR for short-packet communication scenarios.

Note that in existing studies under the Shannon theory, errors can be avoided when the transmission rate is below the Shannon capacity. Outage probability is thus an appropriate performance metric to compare the performance of HDR and FDR \cite{Full_duplex_outage, Full_duplex_adaption}. However, when it comes to short-packet communication scenarios, there always exists a non-zero error probability even when the transmission rate is below Shannon capacity. According to the seminal work given in \cite{FBLtheory1}, the achievable rate in short-packet communications depends on both the blocklength and the desired BLER. We are thus motivated to adopt the BLER as a metric to evaluate and compare the performance of HDR and FDR. Furthermore, the Shannon bound considered in the existing studies only involves a logarithm function and depends on the received SNR. In this sense, the comparison between HDR and FDR for conventional long-packet communications conducted in \cite{Full_duplex_outage, Full_duplex_adaption} thus focused only on the received SNR and coding rate. In contrast, for the considered short-packet communications, we need to jointly consider the received SNR, coding rate, the blocklength and the BLER to properly compare these two duplex modes. Moreover, the new rate bound for short-packet communications has a more complex structure and involves a complicated $Q$-function. As such, the performance analysis and optimization requires more complex mathematical manipulations and new approximation methods.

The main contribution of this paper is summarized as follows. We derive approximate closed-form expressions of average BLER for FDR and HDR under the finite blocklength regime. In order to gain further insights, we then derive simple asymptotic expressions of average BLER at high signal-to-noise ratio (SNR) for both systems. Based on the derived asymptotic expressions, we optimize the transmit powers of source and relay under individual peak power constraints and analytically compare the optimal performance of FDR and HDR. We obtain an expression for a critical BLER that can be used to choose the optimal duplex mode for short-packet communication scenarios. Simulation results are provided to verify the correctness of our theoretical analysis.

\textbf{\emph{Notation}}: Throughout this paper, we use $f_{X}(x)$ and $F_{X}(x)$ to denote the probability density function (PDF) and cumulative distribution function (CDF) of a random variable $X$. $\mathbb{E}\left\{ {\cdot} \right\}$ represents the expectation operator. $Q\left( x \right) = \int_x^\infty  {{1 \over {\sqrt {2\pi } }}{e^{ - {{{t^2}} \over 2}}}dt} $ is the Q-function. ${\rm Ei}\left(x\right)$ is the exponential integral function \cite[Eq. (8.21)]{Tableofintegral}.

\section{System Model}
We consider a dual-hop relay system consisting of one source $S$, one destination $D$ and one decode-and-forward (DF) relay $R$. The relay can work in either full-duplex relaying (FDR) or half-duplex relaying (HDR) modes. The source and destination are equipped with single antenna and the relay is implemented with two isolated receive and transmit antennas. The isolation of antennas at the relay can mitigate the line-of-sight (LoS) component of loop interference (or ``echo interference") caused by receiving and transmitting information simultaneously when operating in the full-duplex mode. However, the loop interference cannot be eliminated completely due to multipath scattering \cite{Full_duplex_outage,Full_duplex_adaption}. We thus consider the existence of loop interference link $R-R$ for FDR (i.e., residual interference after cancellation). Besides, we investigate a short-packet communication scenario and each transmission block of the considered system has a length of $m$ channel uses (c.u.). The links $S-R$, $R-D$, and $R-R$ are assumed to suffer from Rayleigh fading with average power gains $\Omega_{S,R}$, $\Omega_{R,D}$, and $\Omega_{R,R}$, respectively. We assume quasi-static fading channels for which the fading coefficients remain constant during each transmission block and change independently from one block to another. In the following, we give the SNRs at relay and destination for FDR and HDR.

%\subsection{Full-Duplex Relaying (FDR)}
In FDR, the source transmits information to the relay during each whole transmission block. At the same time, the relay forwards the received signal from source to destination. We denote by $P_S$ the transmit power of the source and $P_R$ the transmit power of the relay. We use $N_R$ and $N_D$ to denote the variance of additive white Gaussian noise (AWGN) at $R$ and $D$, respectively. For notation simplicity, we define the average SNR for the considered three links as ${\overline\gamma  _{S,R}} = {{{P_S}\Omega_{S,R}} \over {{N_R}}}$, ${\overline \gamma  _{R,D}} = {{{P_R}\Omega_{R,D}} \over {{N_D}}}$ and ${\overline \gamma  _{R,R}} = {{{P_R}\Omega_{R,R}} \over {{N_R}}}$. According to \cite{Full_duplex_outage,Full_duplex_adaption}, the SNRs at $R$ and $D$ for FDR can be written as ${\gamma _{R}^F} = {{{{\bar \gamma }_{S,R}}X} \over {{{\bar \gamma }_{R,R}}Y + 1}}$, ${\gamma _{D}^F} = {{{\bar \gamma }_{R,D}}Z}$, respectively, where $X$, $Y$ and $Z$ are exponentially distributed random variables with unit power gain.

%\subsection{Half-Duplex Relaying (HDR)}
In HDR, each transmission block is divided into two time slots with equal length. During the first time slot, the source transmits information to the relay. In the second time slot, the relay decodes and forwards the received signal to the destination. In HDR, the loop interference at the relay is completely avoided by receiving and transmitting information separately. The received SNRs at $R$ and $D$ for the HDR system are given by $\gamma_{R}^H = {{{\bar \gamma }_{S,R}}X}$ and $\gamma_{D}^H = {{{\bar \gamma }_{R,D}}Y}$, respectively.

\section{Performance Analysis and Comparison}
In this section, we derive approximate closed-form expressions of BLER for both FDR and HDR under the finite blocklength regime. In order to gain further insights, we then derive asymptotic expressions for BLER at high SNR range and compare these two systems analytically. For a fair comparison, we assume that $S$ transmits $\sigma$ bits of information to $D$ over $m$ channel uses during each packet transmission. Note that in the following analysis of BLER, HDR and FDR have different distributions of end-to-end SNR. Besides, the blocklength and coding rate for each hop is different for HDR and FDR. Specifically, the blocklength of each information packet in FDR and HDR systems is given by $m_F=m$ channel uses and $m_H=m/2$ channel uses, respectively, the coding rate for FDR and HDR is given by ${r_F} = {\sigma  \over {{m}}}$ and ${r_H} = {2\sigma  \over {{m}}}$, respectively.

\subsection{Full-Duplex Relaying (FDR)}
We first study the BLER of the FDR system. Considering a decode-and-forward protocol implemented at the relay, errors can occur from the following two events: relay detects an error and relay decodes the received information correctly but destination detects an error. Let $\varepsilon_{S,R}^F$ and $\varepsilon_{R,D}^F$ denote the BLER of $S-R$ link and $R-D$ link of the FDR system, the overall BLER can thus be expressed as
\begin{equation}\label{eFDR}
{\varepsilon _F}={\varepsilon _{S,R}^F} + \left( {1 - {\varepsilon _{S,R}^F}} \right){\varepsilon _{R,D}^F}.
\end{equation}
We now evaluate the terms ${\varepsilon _{S,R}^F}$ and ${\varepsilon _{R,D}^F}$. Specifically, according to \cite[Eq. (59)]{FBLtheory}, when ${m_F}$ is sufficiently large (i.e., ${m_F}>100$), the term ${\varepsilon _{S,R}^F}$ can be tightly approximated as $\varepsilon_{S,R}^F \approx \mathbb {E} \left\{Q\left( {{{C\left( {{\gamma _{R}^F}} \right) - {r_{F}}} \over {\sqrt {V\left( {{\gamma _{R}^F}} \right)/{m_F}} }}} \right) \right\}$, where the expectation is over the received SNR ${\gamma _{R}^F}$, $C\left( {{\gamma _{R}^F}} \right) = {\log _2}\left( {1 + {\gamma _{R}^F}} \right)$ is the Shannon capacity and $V\left( {{\gamma _{R}^F}} \right) = \left( {1 - {1 \over {{{\left( {1 + {\gamma _{R}^F}} \right)}^2}}}} \right){\left( {{{\log }_2}e} \right)^2}$ is the channel dispersion which measures the stochastic variability of the channel relative to a deterministic channel with the same capacity \cite{FBLtheory1}. It is hard to characterize $\varepsilon_{S,R}^F$ in a closed-form due to the complicated $Q$-function and we are thus motivated to use a linear approximation of $Q\left( {{{C\left( {{\gamma _{R}^F}} \right) - {r_F}} \over {\sqrt {V\left( {{\gamma _{R}^F}} \right)/{m_F}} }}} \right) \approx \Xi\left({\gamma _{R}^F}\right)$ given by \cite{Qapproximation,ultrareliable}
\begin{equation}\label{approximation}
\Xi \left( {\gamma _{R}} \right) = \left\{ {
\begin{matrix}
\begin{split}
   &1, \quad \quad \quad \quad \quad\quad\quad\quad\quad\quad\quad\quad {\gamma _{R}^F} \le  \zeta_{F} \\
   &{{1 \over 2}} -{\vartheta_{F} \sqrt {{m_{F}}}}\left( {{\gamma _{R}^F} - \theta_{F} } \right), \quad   \zeta_{F}   < {\gamma _{R}^F} < \xi_{F}   \\
  & 0,  \quad \quad \quad\quad\quad\quad\quad\quad\quad\quad\quad\quad {\gamma _{R}^F} \ge \xi_{F} \\
\end{split}
 \end{matrix} } \right.,
\end{equation}
where $\vartheta_{F} = {1 \over {2\pi \sqrt {{2^{2{r_{F}}}} - 1} }}$, $\theta_{F} = {2^{{r_{F}}}} - 1$, $\zeta_{F} = \theta_{F}  - {1 \over {2\vartheta_{F} \sqrt {{m_{F}}} }}$ and $\xi_{F} =\theta_{F}  + {1 \over {2\vartheta_{F} \sqrt {{m_{F}}} }} $.
With the above approximation, ${\varepsilon _{S,R}^F}$ can be evaluated as
\begin{equation}\label{app1}
{\varepsilon _{S,R}^F} \approx \int_0^\infty  {\Xi \left( x \right)} {f_{{\gamma _{R}^F}}}\left( x \right)dx = {\vartheta _F}\sqrt {m_F}\int_{{\zeta _F}}^{{\xi _F}} {{F_{{\gamma _{R}^F}}}\left( x \right)}dx.
\end{equation}

With the CDF of ${\gamma _{R}^F}$ given in \cite[Eq. (6)]{Full_duplex_outage} and the integral formula \cite[Eq. (3.352-1)]{Tableofintegral}, the term ${\varepsilon _{S,R}^F}$ can be evaluated from (\ref{app1}) as
\begin{equation}\label{e1}
\begin{split}
{\varepsilon _{S,R}^F} &\approx 1{\rm{ - }}{\vartheta _F}\sqrt {m_F}{{{{\bar \gamma }_{S,R}}} \over {{{\bar \gamma }_{R,R}}}}\exp \left( {{1 \over {{{\bar \gamma }_{R,R}}}}} \right)\times \\
&\quad \left[ {{\rm Ei}\left( { - {{{\xi _F}} \over {{{\bar \gamma }_{S,R}}}} - {1 \over {{{\bar \gamma }_{R,R}}}}} \right) - {\rm Ei}\left( { - {{{\zeta _F}} \over {{{\bar \gamma }_{S,R}}}} - {1 \over {{{\bar \gamma }_{R,R}}}}} \right)} \right].
\end{split}
\end{equation}

We can also evaluate the term ${\varepsilon _{R,D}^F}$ similarly. By substituting ${\varepsilon _{S,R}^F}$, ${\varepsilon _{R,D}^F}$ into (\ref{eFDR}), the overall BLER for FDR system can be written in (\ref{eFDR1}) on top of the next page.
\begin{figure*}[!t]
% ensure that we have normalsize text
%\normalsize
%% Store the current equation number.
%\setcounter{mytempeqncnt}{\value{equation}}
%% Set the equation number to one less than the one
%% desired for the first equation here.
%% The value here will have to changed if equations
%% are added or removed prior to the place these
%% equations are referenced in the main text.
%\setcounter{equation}{30}.
\begin{equation}\label{eFDR1}
\begin{split}
{\varepsilon _F} & \approx 1{\rm{ - }}{\vartheta _F}^2{m_F}{{{{\bar \gamma }_{S,R}}{{\bar \gamma }_{R,D}}} \over {{{\bar \gamma }_{R,R}}}}{e^{{1 \over {{{\bar \gamma }_{R,R}}}}}}\left[ {{\rm Ei}\left( { - {{{\xi _F}} \over {{{\bar \gamma }_{S,R}}}} - {1 \over {{{\bar \gamma }_{R,R}}}}} \right) - {\rm Ei}\left( { - {{{\zeta _F}} \over {{{\bar \gamma }_{S,R}}}} - {1 \over {{{\bar \gamma }_{R,R}}}}} \right)} \right]\left[ {\exp \left( { - {{{\zeta _F}} \over {{{\bar \gamma }_{R,D}}}}} \right) - \exp \left( { - {{{\xi _F}} \over {{{\bar \gamma }_{R,D}}}}} \right)} \right].
\end{split}
\end{equation}
% Restore the current equation number.
%\setcounter{equation}{\value{mytempeqncnt}}
% IEEE uses as a separator
\hrulefill
% The spacer can be tweaked to stop underfull vboxes.
\vspace*{4pt}
\end{figure*}

\subsection{Half-Duplex Relaying (HDR)}
We now consider the HDR system and for the purpose of brevity, we omit the details of the derivation. With a similar method used in the previous subsection, we can attain the overall BLER for the HDR system approximated by
\begin{equation}\label{eHDR}
\begin{split}
{\varepsilon _H} &\approx 1{\rm{ - }}{\vartheta _H}^2{m_H}{{\bar \gamma }_{S,R}}{{\bar \gamma }_{R,D}}\left( {{e^{ - {{{\zeta _H}} \over {{{\bar \gamma }_{S,R}}}}}} - {e^{ - {{{\xi _H}} \over {{{\bar \gamma }_{S,R}}}}}}} \right)\\
&\quad \times \left( {{e^{ - {{{\zeta _H}} \over {{{\bar \gamma }_{R,D}}}}}} - {e^{ - {{{\xi _H}} \over {{{\bar \gamma }_{R,D}}}}}}} \right),
\end{split}
\end{equation}
where $m_H = {m \over 2}$ is the blocklength for HDR system, ${r_H} = {2\sigma  \over {{m}}}$ is the coding rate for HDR and the parameters $\zeta_{H}$, $\xi_{H}$ are given by $\zeta_{H} = \theta_{H}  - {1 \over {2\vartheta_{H} \sqrt {{m_{H}}} }}$, $\xi_{H} =\theta_{H}  + {1 \over {2\vartheta_{H} \sqrt {{m_{H}}} }} $, where $\vartheta_{H} = {1 \over {2\pi \sqrt {{2^{2{r_{H}}}} - 1} }}$ and $\theta_{H} = {2^{{r_{H}}}} - 1$.

Till now, we have derived closed-form expressions for the BLER of both FDR and HDR, respectively. However, due to the complicated structure of the expressions, we cannot gain further insights in terms of the effect of various system parameters on the system performance. We are thus motivated to conduct asymptotic analysis by deriving simple expressions at high SNR in the following subsections.

\subsection{Asymptotic Performance at High SNR}
At high SNR, the received SINR at $R$ in FDR can be expressed as ${\gamma _{R}^F} \approx {{{{\bar \gamma }_{S,R}}X} \over {{{\bar \gamma }_{R,R}}Y }}$. The CDF of $\gamma_R^F$ can be approximated~as
\begin{equation}\label{highSNRCDF}
\begin{split}
{F_{{\gamma _R^F}}}\left( x \right)&=\int_0^\infty  {\left[ {1{\rm{ - }}{\exp\left({ - {{xt} \over {{{\bar \gamma }_{S,R}}}}}\right)}} \right]} {1 \over {{{\bar \gamma }_{R,R}}}}\exp \left( { - {t \over {{{\bar \gamma }_{R,R}}}}} \right)dt\\
& \approx {x \over {{{\bar \gamma }_{S,R}}{{\bar \gamma }_{R,R}}}}\int_0^\infty  t \exp \left( { - {t \over {{{\bar \gamma }_{R,R}}}}} \right)dt\\
& = {{{{\bar \gamma }_{R,R}}} \over {{{\bar \gamma }_{S,R}}}}x,
\end{split}
\end{equation}
where we apply approximation $1 - \exp \left( { - {{xt} \over {{{\bar \gamma }_{S,R}}}}} \right) \approx {{xt} \over {{{\bar \gamma }_{S,R}}}}$ in the above derivation and the last integral can be solved by using \cite[Eq. (3.351-3)]{Tableofintegral}. On the other hand, the CDF of the received SNR at $D$ can be approximated as ${F_{{\gamma _D^F}}}\left( x \right) = 1 - \exp \left( { - {x \over {{{\bar \gamma }_{R,D}}}}} \right) \approx {x \over {{{\bar \gamma }_{R,D}}}}$. With the above two simplified CDFs and the expression given in (\ref{app1}), the BLER for the FDR system at high SNR can be derived as
\begin{equation}\label{highSNRFDR}
\begin{split}
\varepsilon _F^\infty  &= {\varepsilon _{S,R}^F} + {\varepsilon _{R,D}^F}-{\varepsilon _{S,R}^F}{\varepsilon _{R,D}^F} \\
& \approx {\varepsilon _{S,R}^F} + {\varepsilon _{R,D}^F}\\
& \approx \left( {{{{{\bar \gamma }_{R,R}}} \over {{{\bar \gamma }_{S,R}}}}+{{1} \over {{{\bar \gamma }_{R,D}}}}} \right)\left( {{2^{{\sigma  \over m}}} - 1} \right).
\end{split}
\end{equation}

\begin{remark}
\iffalse
From (\ref{highSNRFDR}), we can see that the BLER of the FDR system decreases as the source transmit power $P_S$ increases. However, the increase of relay transmit power $P_R$ places two opposite impacts on the overall BLER. On one hand, it increases the BLER by increasing the term ${{\bar \gamma }_{R,R}}$, on the other hand, it decreases the BLER by increasing the term ${{{\bar \gamma }_{R,D}}}$. This observation coincide well with the conventional FDR systems.
\fi

We now optimize the source transmit power and relay transmit power by assuming that they are subject to individual peak power constraints given by $P_S, P_R \le P_C$. Obviously, the optimal source transmit power is given by $P_S^* = P_C$. Substituting $P_S^*= P_C$ into (\ref{highSNRFDR}), taking the derivative with respect to $P_R$ {and considering the power constraint,} we can obtain the optimal value of relay transmit power as $P_R^* = \min\left\{P_C, \sqrt {{{{P_C}{\Omega _{S,R}}{N_D}} \over {{\Omega _{R,R}}{\Omega _{R,D}}}}} \right\}$. We surprisingly find out that the optimal values $P_S^*$ and $P_R^*$ at high SNR for a FDR under the finite blocklength regime is independent of the blocklength $m$. This is due to the approximation methods adopted in the asymptotic analysis of FDR. Besides, $P_R^*$ is proportional to $\Omega_{S,R}$ but inversely proportional to $\Omega_{R,D}$, $\Omega_{R,R}$.

\iffalse
With the optimal transmit powers, the minimum BLER for a FDR system at high SNR is given by
\begin{equation}\label{optimalFDR}
\hat \varepsilon _F^\infty \approx \left( {{{{\min \left\{ {{P_C},\sqrt {{{{P_C}{\Omega _{S,R}}{N_D}} \over {{\Omega _{R,R}}{\Omega _{R,D}}}}} } \right\}}{\Omega _{R,R}}} \over {P_C{\Omega _{S,R}}}} + {{{N_D}} \over {\min \left\{ {{P_C},\sqrt {{{{P_C}{\Omega _{S,R}}{N_D}} \over {{\Omega _{R,R}}{\Omega _{R,D}}}}} } \right\}{\Omega _{R,D}}}}} \right)\left( {{2^{{\sigma  \over m}}} - 1} \right).
\end{equation}
\fi
\end{remark}

Similarly, the BLER for a HDR system at high SNR range can be asymptotically expressed~as
\begin{equation}\label{optimalHDR}
\hat \varepsilon _H^\infty   \approx \left( {{{{N_R}} \over {P_S{\Omega _{S,R}}}} + {{{N_D}} \over {P_R{\Omega _{R,D}}}}} \right)\left( {{2^{{{2\sigma } \over m}}} - 1} \right).
\end{equation}

For individual constraints $P_R, P_S \le P_C$, the optimal transmit powers of HDR systems are $P_S^*= P_R^* = P_C$.

\subsection{Performance Comparison}
In the following, we analytically compare the performance of HDR and FDR with optimal transmit powers at source and relay for short-packet communications. Specifically, we define $\delta_F$ and $\delta_H$ as the minimum blocklength (i.e., delay) for the considered FDR and HDR systems under a given BLER constraint $\varepsilon$, respectively. Based on the results given in (\ref{highSNRFDR}) and (\ref{optimalHDR}), the minimum delay, measured by channel uses, can be expressed as $\delta_F = {\sigma  \over {{{\log }_2}\left( {{\varepsilon \over A} + 1} \right)}}$ and $\delta_H = {{2\sigma } \over {{{\log }_2}\left( {{\varepsilon \over B} + 1} \right)}}$, where $A={{{{\min \left\{ {{P_C},\sqrt {{{{P_C}{\Omega _{S,R}}{N_D}} \over {{\Omega _{R,R}}{\Omega _{R,D}}}}} } \right\}}{\Omega _{R,R}}} \over {P_C{\Omega _{S,R}}}} + {{{N_D}} \over {\min \left\{ {{P_C},\sqrt {{{{P_C}{\Omega _{S,R}}{N_D}} \over {{\Omega _{R,R}}{\Omega _{R,D}}}}} } \right\}{\Omega _{R,D}}}}}$ and $B={1 \over {{P_C}}}\left( {{{{N_R}} \over {{\Omega _{S,R}}}} + {{{N_D}} \over {{\Omega _{R,D}}}}} \right)$. We now define the term $\Delta\delta=\delta_F-\delta_H$. We then have HDR outperforms FDR when $\Delta\delta>0$ since HDR system requires a shorter blocklength for a common BLER constraint and vice versa. After some manipulation, the term $\Delta\delta$ can be evaluated as
\begin{equation}\label{M}
\begin{split}
\Delta\delta = {{\sigma {{\log }_2}\left[ {1 + {\varepsilon  \over A}\left( {{A \over B} - {\varepsilon  \over A} - 2} \right)} \right]} \over {{{\log }_2}\left( {A + 1} \right){{\log }_2}\left( {B + 1} \right)}}.\\
\end{split}
\end{equation}

\begin{remark}
From (\ref{M}), we immediately have $\Delta\delta<0$ when the term ${{A \over B} - {\varepsilon  \over A} - 2}<0$ since ${\varepsilon  \over A}>0$. Moreover, ${{A \over B} - {\varepsilon  \over A} - 2}<0$ always holds for any given value of $\varepsilon$ when $A<2B$. This means that when $A<2B$, FDR system always achieves shorter delay than HDR system for any BLER requirement. We next investigate the case when $A \ge 2B$. To this end, we define a critical value of the required BLER, denoted by $\varepsilon^*$, which satisfies $\Delta\delta=0$. We can easily attain that ${\varepsilon ^{\rm{*}}} = {A \over B}\left( {A{\rm{ - }}2B} \right)$. We can also verify that $\Delta\delta$ is a decreasing function of $\varepsilon$. We thus can deduce that for any given BLER requirement $\varepsilon > \varepsilon^*$, we have $\Delta\delta<0$ and FDR outperforms HDR in terms of shorter delay to achieve the same reliability. On the other hand, for any $\varepsilon < \varepsilon^*$, we have $\Delta\delta>0$ and HDR system outperforms FDR system. This also indicates that FDR is more appealing to a system with looser requirements on BLER. This is understandable since FDR system suffers from loop interference, while HDR system avoids such interference completely. The critical value $\varepsilon^*$ can be expressed as
\begin{equation}\label{criticale}
{\varepsilon ^{\rm{*}}}{\rm{ = }}\left\{ {
\begin{matrix}
   {{A \over B}\left( {A{\rm{ - }}2B} \right), \quad \text{if} \quad A \ge 2B}  \\
   {\text{Not Exist}}, \quad \text{Otherwise}  \\
\end{matrix} } \right..
\end{equation}
\iffalse{\color{blue}From the above analysis, the relaying duplex mode selection criteria under short-packet communication scenario is given by
\begin{equation}
\left\{
\begin{matrix}
   {\varepsilon  < {\varepsilon ^*} \to \text{HDR}}  \\
   {\varepsilon  \ge {\varepsilon ^*} \to \text{FDR}}  \\
\end{matrix}  \right.,
\end{equation}
where $\varepsilon$ is the target system BLER and the critical value $\varepsilon^*$ is defined as
\begin{equation}\label{criticale}
{\varepsilon ^{\rm{*}}}{\rm{ = }}\left\{ {
\begin{matrix}
   {{A \over B}\left( {A{\rm{ - }}2B} \right), \quad \text{if} \quad A \ge 2B}  \\
   {\text{Not Exist}}, \quad \text{Otherwise}  \\
\end{matrix} } \right..
\end{equation}}
\fi
\end{remark}

\section{Numerical Results and Discussions}
In this section, we present some numerical results to validate our theoretical analysis  performed in this letter. Unless specified otherwise, we set the average channel power gains of the $S$-$R$ and $R$-$D$ links as $\Omega_{S,R}=-80$dB, $\Omega_{R,D}=-85$dB, respectively. The variance of AWGN suffered at $R$ and $D$ are set as $N_R=N_D=-90$dBm.

\begin{figure}
\centering
  {\scalebox{0.40}{\includegraphics {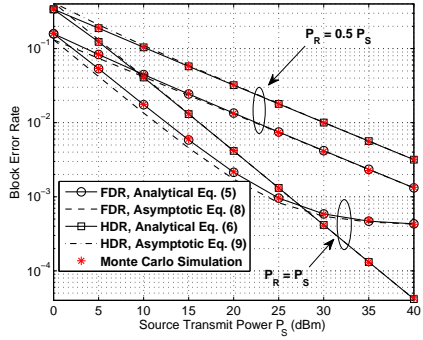}}}
\caption{Analytical BLER and asymptotic BLER versus $P_S$ for different $P_R$, where $\Omega_{R,R}=-110$dB, $\sigma = 256$ bits and $m=512$ channel uses.
\label{montsimu}}
\end{figure}

We first compare the analytical expressions of average BLER for FDR and HDR systems with the corresponding Monte Carlo simulations in Fig. \ref{montsimu}. We can first observe from Fig. \ref{montsimu} that the derived analytical expressions match the simulation results very well and the asymptotic expressions approach the simulation results as source transmit power increases. These observations validate our theoretical analysis performed in Sec. III. Besides, we can see that HDR and FDR systems can outperform each other depending on different system setups, and FDR is preferable for a system with relatively lower relay transmit power. This is because that an increase of relay transmit power can increase the loop interference suffered in the FDR mode. The lower the relay transmit power, the weaker the influence of loop interference. As the analytical results agree well with the simulation, we will only plot the analytical values in the following figures.

\begin{figure}
\centering
  {\scalebox{0.40}{\includegraphics {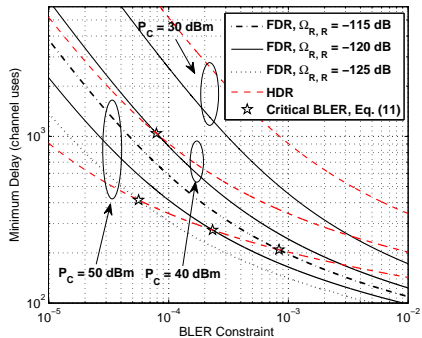}}}
\caption{Minimum delay versus BLER constraint for different transmit power constraints, where $\sigma = 800$ bits.
\label{eplot}}
\end{figure}

{We now compare the latency performance (in channel uses) of HDR and FDR in Fig. \ref{eplot} with their optimal settings.} From Fig. \ref{eplot}, we first observe that for the case $P_C =30$dBm and $\Omega_{R,R} = -120$dB, FDR always outperforms HDR for any given value of BLER requirement. This coresponds to the case $A \le 2B$ given in Remark 2 and the critical BLER does not exist. {For all other simulated cases, there exists a critical value of BLER and the performance superiority of HDR and FDR is reversed before and after this critical BLER.} Besides, we can see that the critical BLER increases as the power constraint grows, and FDR is thus more suitable for a system with relatively lower power budget. This is understandable as unlike the HDR which uses the maximum transmit power for source and relay to send information, FDR needs to balance the relay transmit power with loop interference and the optimal relay transmit power may not be the maximum value. Finally, we can observe that the critical value of BLER shifts to the right as the average power of loop interference link $\Omega_{R,R}$ decreases. This observation reveals that FDR is preferable for a system with a weaker loop interference link which coincides well with the conventional FDR systems.

\section{Conclusions}
In this letter, we revisited full-duplex relaying (FDR) and half-duplex relaying (HDR) under the finite blocklength regime. Specifically, we first derived approximate closed-form expressions of average block-error rate (BLER) for both FDR and HDR. In order to gain further insights and optimize the transmit powers of source and relay, we then characterized simple asymptotic expressions in terms of BLER at high SNR for both schemes. To compare the FDR and HDR, we defined and obtained a critical BLER which can be used to determine the performance superiority of FDR and HDR in terms of lower delay under a given BLER requirement. Our results discovered that FDR is more preferable for a system with relatively lower transmit power constraint, less stringent BLER requirement and stronger loop interference mitigation.

%\begin{appendix}
%\input{appendix}
%%\input{appendix_B}
%%\input{appendix_C}
%%\input{appendix_D}
%\end{appendix}

\ifCLASSOPTIONcaptionsoff
  \newpage
\fi

%\section{Acknowledgement}
%The authors would like to thank %the anonymous reviewers for their valuable comments and suggestions, which improved the quality of the paper. The authors also thank

\bibliographystyle{IEEEtran}
\bibliography{References}

%\newpage
%

\end{document}